# Hosting Capacity Assessment and Enhancement for Edge Data Centers in Active Distribution Networks

Linhan Fang
Department of Electrical and Computer Engineering
University of Houston
Houston, TX, USA
lfang7@uh.edu

Xingpeng Li
Department of Electrical and Computer Engineering
University of Houston
Houston, TX, USA
xli82@uh.edu

***Abstract*—With the increasing demand for edge computing and AI-driven workloads, integrating small and medium-sized edge data centers into distribution networks has become increasingly important. This paper investigates the hosting capacity of distribution networks for data center integration and identifies the key physical mechanisms that limit the maximum allowable data center load. The baseline analysis shows that data center hosting capacity varies significantly across candidate buses due to network topology and electrical distance. Three dominant limiting mechanisms are identified: current-constrained locations, voltage-constrained locations, and mixed-constrained locations where both current loading and voltage deviation jointly affect hosting capacity. To increase the hosting capacity, this study evaluates multiple flexible resources, including battery energy storage systems (BESS), dispatchable distributed generators (DDG), and static synchronous compensators (STATCOM). Numerical results demonstrate that these resources provide complementary benefits through active power support, sustained local generation, and reactive power compensation, effectively expanding data center hosting capacity in distribution systems.**

## I. INTRODUCTION

With the rapid proliferation of artificial intelligence and advanced data analytics, the deployment of data center (DC) infrastructure has experienced significant acceleration [1]. To facilitate low-latency data acquisition and real-time processing, there is an increasing trend toward integrating small-to-medium-scale edge data centers directly into distribution networks. This paradigm shift requires a comprehensive evaluation of the hosting capacity (HC) and energy integration capabilities of the distribution network. Specifically, the network must accommodate the intensive power demands of DCs while ensuring operational stability and reliability.

The hosting capacity serves as a fundamental metric for assessing the security and stability of a distribution network. It quantifies the maximum penetration level of data center loads that the infrastructure can sustain without violating technical constraints [2]-[3]. Consequently, conducting a rigorous HC analysis is a critical prerequisite during the initial planning stages of data center integration. Data centers are characterized by high power density, which significantly exceeds the demand profiles of conventional residential or commercial loads. To enhance the network's ability to support these concentrated loads, the coordination of multiple power sources is essential [4]-[5]. If data centers are treated merely as passive loads, the total permissible capacity is strictly constrained by line thermal limits and terminal voltage limits.

However, when various types of distributed energy resources (DERs) are considered, the hosting capacity becomes a dynamic boundary influenced by system topology and available flexibility [6]-[7]. For instance, the strategic deployment of dispatchable distributed generators (DDG) and battery energy storage systems (BESS) can provide localized support, thereby mitigating the peak power demand imposed on the upstream feeder sections and transmission networks. Furthermore, the implementation of reactive power compensation at the distribution feeders can effectively improve the voltage profile, further expanding the margin for data center integration.

Existing HC studies have mainly focused on distributed generation, photovoltaic generation, and electric vehicle integration. A streamlined HC assessment method was proposed in [8] to evaluate distribution feeders without requiring detailed interconnection studies for every scenario. EV-related HC studies further showed that load growth can create new constraints in distribution and microgrid operation, requiring coordinated control strategies [9]-[10]. Other studies have investigated the role of network reconfiguration in increasing distributed generation HC [11], as well as the coordinated operation of OLTC and SVC devices to improve HC under voltage and operating constraints [12]. These studies demonstrate that HC is strongly location-dependent and is commonly limited by voltage deviation, line ampacity, transformer loading, and control device capability.

Several enhancement methods have also been studied to increase the ability of distribution networks to accommodate new resources. BESS management and curtailment-based strategies have been used to mitigate voltage and thermal violations under high DER penetration [13]. Operational resources and demand response have also been considered to improve distribution system flexibility and resilience [14]. In addition, STATCOM-based reactive power compensation has been widely recognized as an effective approach for dynamic voltage support and power quality improvement [15]. Nevertheless, limited attention has been given to DC-oriented HC assessment, where concentrated active power demand, sustained local generation, storage support, and reactive compensation may affect HC through different mechanisms.

## II. METHODOLOGY

This section proposes an hourly-resolution distribution-flow optimization model to analyze the HC of a single DC installed in a distribution network, and the proposed model tests the impact of installing BESS, DDG, and STATCOM on HC.

### *A. Baseline Data Center Hosting Capacity Model*

The objective of the baseline model is to maximize the HC for the DC on each potential bus. For each simulation case, one candidate bus $b$ is fixed as the DC installation bus. The DC active and reactive loads are defined in (2):

$$\delta_{i,b} = \begin{cases} 1, & i = b \\ 0, & i \neq b \end{cases} \tag{1}$$

$$\begin{cases} P_{\mathrm{DC},i} = \delta_{i,b} H_b^{\mathrm{ori}} \\ Q_{\mathrm{DC},i} = \delta_{i,b} H_b^{\mathrm{ori}} \tan\theta \end{cases} \tag{2}$$

$$\text{objective} = \max H_b \tag{3}$$

where $H_b^{\mathrm{ori}}$ denotes the original HC when the DC is installed at the fixed bus $b$. $P_{\mathrm{DC},i}$ and $Q_{\mathrm{DC},i}$ are non-negative DC load reflecting the constant power consumption characteristics.

Nodal active and reactive power balance equations at non-substation and substation buses are expressed as (4)-(5).

$$\begin{cases} \sum_{k\in\mathcal{U}(i)} (P_{ki}^t - r_{ki}\cdot l_{ki}^t) = \sum_{j\in\mathcal{D}(i)} P_{ij}^t + P_{\mathrm{load},i}^t + P_{\mathrm{DC},i} \\ \sum_{k\in\mathcal{U}(i)} (Q_{ki}^t - x_{ki}\cdot l_{ki}^t) = \sum_{j\in\mathcal{D}(i)} Q_{ij}^t + Q_{\mathrm{load},i}^t + Q_{\mathrm{DC},i} \end{cases} \tag{4}$$
$$\forall i \in \mathcal{N}, \forall t \in \mathcal{T}$$

$$\begin{cases} \sum_{k\in\mathcal{U}(g)} (P_{kg}^t - r_{kg}\cdot l_{kg}^t) = -P_g^t + \sum_{j\in\mathcal{D}(g)} P_{gj}^t + P_{\mathrm{load},g}^t \\ \sum_{k\in\mathcal{U}(g)} (Q_{kg}^t - x_{kg}\cdot l_{kg}^t) = -Q_g^t + \sum_{j\in\mathcal{D}(g)} Q_{gj}^t + Q_{\mathrm{load},g}^t \\ \forall g \in \mathcal{G}, \forall t \in \mathcal{T} \end{cases} \tag{5}$$

where for each node $i \in \mathcal{N}$, $\mathcal{D}(i)$ and $\mathcal{U}(i)$ represent the sets of its downstream and upstream nodes, respectively.

Voltage drop equation can be obtained as (6) and the constraint between current and power is defined in (7) where $\mathcal{L}$ denotes the branch information:

$$v_i^t - v_j^t = 2\left(r_{ij}P_{ij}^t + x_{ij}Q_{ij}^t\right) - \left(r_{ij}^2 + x_{ij}^2\right) l_{ij}^t$$
$$, \forall (i,j) \in \mathcal{L}, \forall t \in \mathcal{T} \tag{6}$$

$$v_i^t \cdot l_{ij}^t \geq (P_{ij}^t)^2 + (Q_{ij}^t)^2, \forall (i,j) \in \mathcal{L}, \forall t \in \mathcal{T} \tag{7}$$

The constraints of voltage and current are shown in (8)-(9), and $v^{\min}$ and $v^{\max}$ are the square of minimum and maximum per-unit voltage, and $I_{ij}^{\lim}$ represents the current rating between node $i$ and $j$.

$$v^{\min} \leq v_i^t \leq v^{\max}, \forall i \in \mathcal{N}, \forall t \in \mathcal{T} \tag{8}$$

$$0 \leq l_{ij}^t \leq (I_{ij}^{\lim})^2, \forall (i,j) \in \mathcal{L}, \forall t \in \mathcal{T} \tag{9}$$

### *B. Flexible Resources Modeling*

As enforced in (10)-(12), $u_{i,t}^{\mathrm{ch}}$ and $u_{i,t}^{\mathrm{dis}}$ are binary variables indicating charging and discharging states of the BESS. Similarly, $u_{i,t}^{\mathrm{inj}}$ and $u_{i,t}^{\mathrm{abs}}$ indicate the reactive power injection and absorption status. These constraints prevent simultaneous charging and discharging of active power, as well as simultaneous reactive power injection and absorption.

$$u_{i,t}^{\mathrm{inj}}, u_{i,t}^{\mathrm{abs}}, u_{i,t}^{\mathrm{ch}}, u_{i,t}^{\mathrm{dis}} \in \{0,1\} \tag{10}$$

$$u_{i,t}^{\mathrm{ch}} + u_{i,t}^{\mathrm{dis}} \leq 1, \forall i \in \mathcal{N}, \forall t \in \mathcal{T} \tag{11}$$

$$u_{i,t}^{\mathrm{inj}} + u_{i,t}^{\mathrm{abs}} \leq 1, \forall i \in \mathcal{N}, \forall t \in \mathcal{T} \tag{12}$$

A proportional power sizing coefficient $\lambda^B$ is employed to bound the BESS allocation. This ensures the deployed BESS size aligns with the local grid strength indicated by the original HC $H_i^{\mathrm{ori}}$ at each bus, avoiding potential violations caused by oversized installations. The rated active power and energy capacity of BESS are defined as $\bar{P}_i$ and $\bar{E}_i$, where $T^B$ denotes the BESS duration and $c^B$ represents the corresponding C-rate.

$$\bar{P}_b = \lambda^B H_b^{\mathrm{ori}} \tag{13}$$

$$\bar{E}_i = T^B \bar{P}_i = \bar{P}_i / c^B, \forall i \in \mathcal{N} \tag{14}$$

Constraints (15)-(18) limit the BESS active and reactive power outputs. The associated binary variables activate the corresponding operating modes.

$$0 \leq P_{i,t}^{\mathrm{BESS_ch}} \leq \bar{P}_i \cdot u_{i,t}^{\mathrm{ch}}, \forall i \in \mathcal{N}, \forall t \in \mathcal{T} \tag{15}$$

$$0 \leq P_{i,t}^{\mathrm{BESS_dis}} \leq \bar{P}_i \cdot u_{i,t}^{\mathrm{dis}}, \forall i \in \mathcal{N}, \forall t \in \mathcal{T} \tag{16}$$

$$0 \leq Q_{i,t}^{\mathrm{BESS_inj}} \leq S_{\mathrm{inv},i} \cdot u_{i,t}^{\mathrm{inj}}, \forall i \in \mathcal{N}, \forall t \in \mathcal{T} \tag{17}$$

$$0 \leq Q_{i,t}^{\mathrm{BESS_abs}} \leq S_{\mathrm{inv},i} \cdot u_{i,t}^{\mathrm{abs}}, \forall i \in \mathcal{N}, \forall t \in \mathcal{T} \tag{18}$$

The inverter capacity of BESS is defined in (19). The relationship between active power, reactive power, and the apparent power of the inverter is given by (20).

$$S_{\mathrm{inv},i} = \kappa_{\mathrm{inv}} \cdot c^B \cdot \bar{E}_i, \forall i \in \mathcal{N} \tag{19}$$

$$\left(P_{i,t}^{\mathrm{BESS_ch}} + P_{i,t}^{\mathrm{BESS_dis}}\right)^2 + \left(Q_{i,t}^{\mathrm{BESS_abs}} + Q_{i,t}^{\mathrm{BESS_inj}}\right)^2 \leq S_{\mathrm{inv},i}^2 \tag{20}$$

The state of charge (SoC) of the BESS depends on its previous state and the charging or discharging actions taken, as shown in (22). To ensure cyclical operation, the model constrains the SoC at the beginning and end time of the optimization period to be equal.

$$SoC^{\min} \cdot \bar{E}_i \leq E_{i,t}^{\mathrm{BESS}} \leq SoC^{\max} \cdot \bar{E}_i, \forall i \in \mathcal{N}, \forall t \in \mathcal{T} \tag{21}$$

$$E_{i,t}^{\mathrm{BESS}} = E_{i,t-1}^{\mathrm{BESS}} + \left(P_{i,t}^{\mathrm{BESS_ch}} \eta_{\mathrm{ch}} - P_{i,t}^{\mathrm{BESS_dis}} / \eta_{\mathrm{dis}}\right) \Delta t, \forall i, t \tag{22}$$

$$E_{i,t_0}^{\mathrm{BESS}} = E_{i,t_{\mathrm{end}}}^{\mathrm{BESS}}, \forall i \in \mathcal{N} \tag{23}$$

Nodal power balance equations for non-substation nodes considering BESS are defined in (24) and (25).

$$\sum_{k\in\mathcal{U}(i)} (P_{ki}^t - r_{ki}\cdot l_{ki}^t) = \sum_{j\in\mathcal{D}(i)} P_{ij}^t + P_{\mathrm{load},i}^t + P_{\mathrm{DC},i}$$
$$+\delta_{i,b}\left(P_{b,t}^{\mathrm{BESS_ch}} - P_{b,t}^{\mathrm{BESS_dis}}\right), \forall i \in \mathcal{N}, \forall t \in \mathcal{T} \tag{24}$$

$$\sum_{k\in\mathcal{U}(i)} (Q_{ki}^t - x_{ki}\cdot l_{ki}^t) = \sum_{j\in\mathcal{D}(i)} Q_{ij}^t + Q_{\mathrm{load},i}^t + Q_{\mathrm{DC},i}$$
$$+\delta_{i,b}\left(Q_{b,t}^{\mathrm{BESS_abs}} - Q_{b,t}^{\mathrm{BESS_inj}}\right), \forall i \in \mathcal{N}, \forall t \in \mathcal{T} \tag{25}$$

The DDG is modeled as an on-site active power support resource co-located with the DC at the selected bus $b$. The rated active power $P_i^{\mathrm{DDG}}$ is determined proportionally to the original hosting capacity $H_i^{\mathrm{ori}}$. The parameter $\lambda^G$ denotes the DDG sizing coefficient.

$$P_b^{\mathrm{DDG}} = \lambda^G H_b^{\mathrm{ori}} \tag{26}$$

$$\sum_{k\in\mathcal{U}(i)} (P_{ki}^t - r_{ki}\cdot l_{ki}^t) = \sum_{j\in\mathcal{D}(i)} P_{ij}^t + P_{\mathrm{load},i}^t + P_{\mathrm{DC},i} - \delta_{i,b} P_b^{\mathrm{DDG}} \tag{27}$$

The STATCOM is modeled as a controllable reactive power compensation device. It is installed either at the same bus as the DC or at the identified weakest voltage bus. The parameter $s$ denotes the selected installation bus, and $\delta_{i,s}$ indicates whether bus $i$ is the selected STATCOM installation bus. A positive $Q_{b,t}^{\mathrm{S}}$ represents reactive power injection from the STATCOM.

$$\sum_{k\in\mathcal{U}(i)} (Q_{ki}^t - x_{ki}\cdot l_{ki}^t) = \sum_{j\in\mathcal{D}(i)} Q_{ij}^t + Q_{\mathrm{load},i}^t + Q_{\mathrm{DC},i} - \delta_{i,s} Q_{s,t}^{\mathrm{S}} \tag{28}$$

## III. BASELINE DC HOSTING CAPACITY RESULTS

This section presents the baseline data center hosting capacity without flexible resource support. For each candidate bus, the maximum installable data center active load is calculated under voltage and branch current constraints. The obtained baseline HC values serve as the reference for evaluating the improvement achieved by BESS, DDG, and STATCOM.

Fig. 1 shows the 24-hour active and reactive load profiles used as the base loading condition. The load exhibits a clear daily variation, with higher active and reactive demand during the daytime period. This profile is used to evaluate whether additional data center load can be accommodated without violating voltage and line current limits.

Fig. 2 presents the baseline data center hosting capacity for different candidate installation buses. The results are grouped by feeder section to highlight the spatial variation of HC. Buses located closer to the substation or on electrically stronger feeder sections generally allow higher data center capacity, while downstream and lateral-end buses show lower HC.

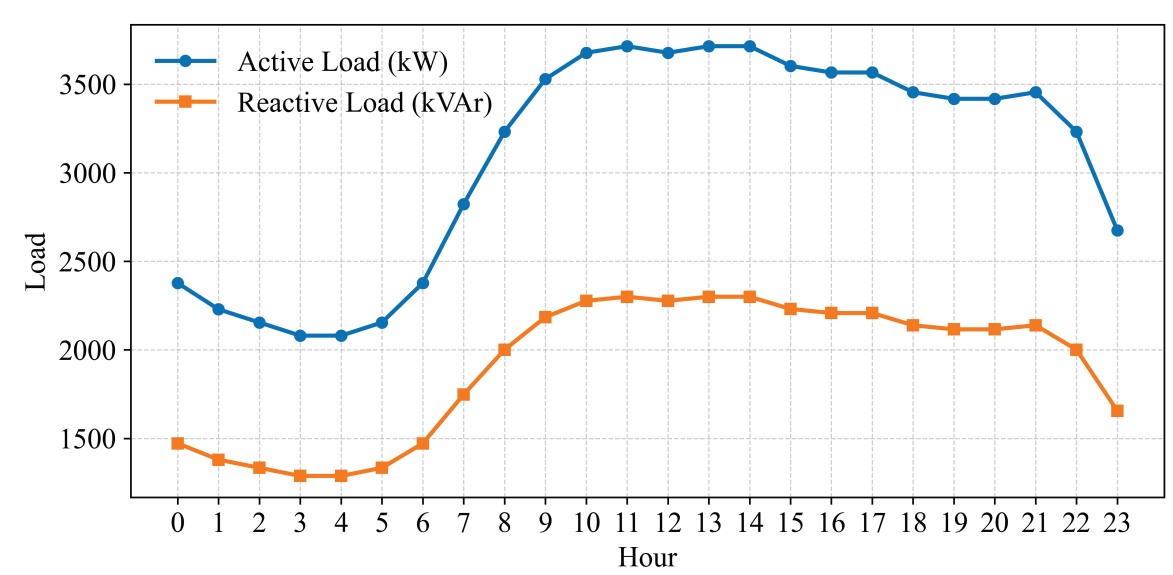

Fig. 1: Daily active and reactive load profiles in IEEE 33-bus system.

Fig. 3 shows the voltage distribution after installing the maximum allowable DC load at each candidate bus. The red regions indicate buses with lower voltage magnitudes, showing where voltage drop becomes a limiting factor for higher HC.

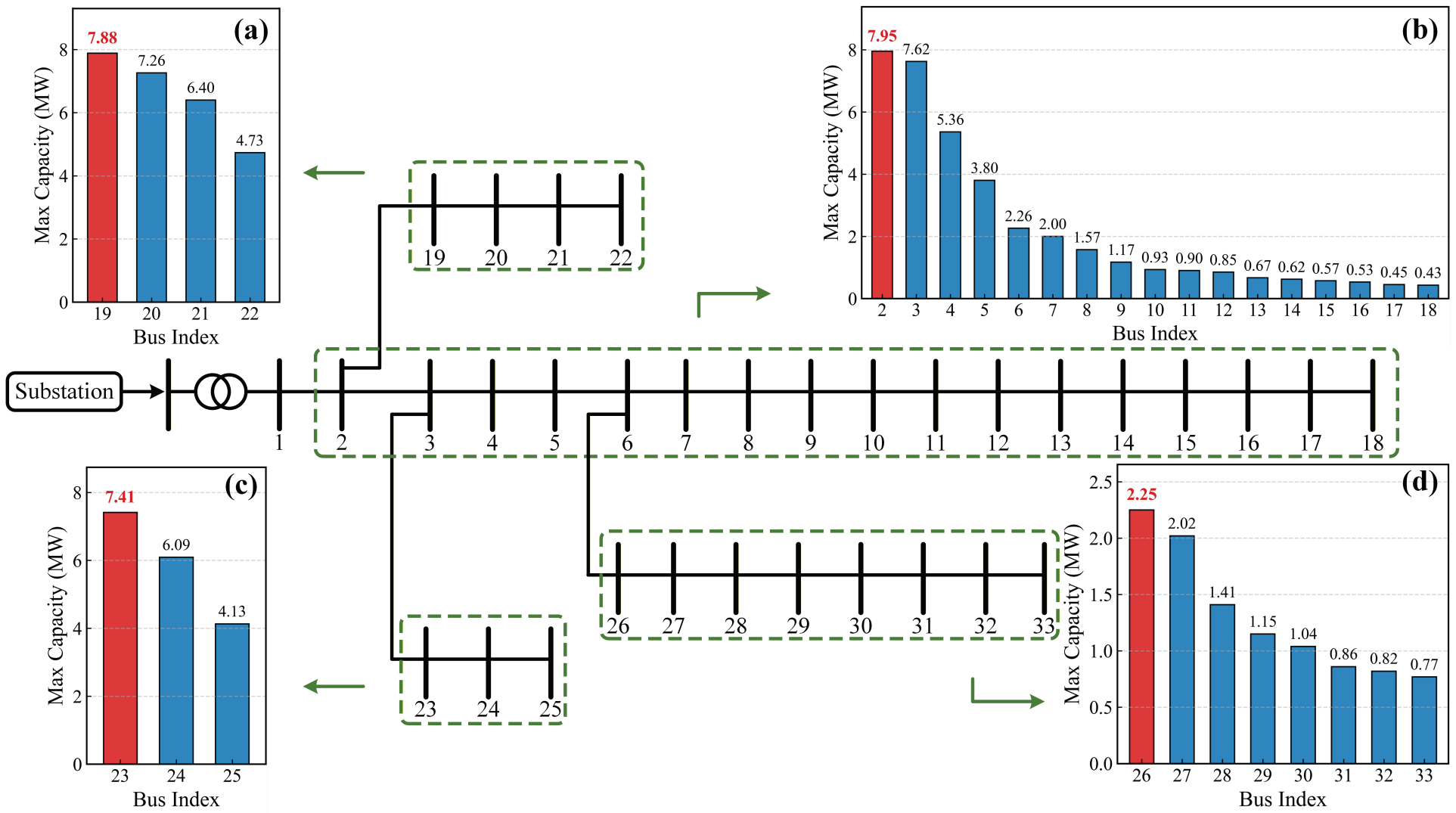

Fig. 2: Baseline HC across candidate buses in different feeder sections: (a) Buses 19-22, (b) Buses 2-18, (c) Buses 23-25, and (d) Buses 26-33.

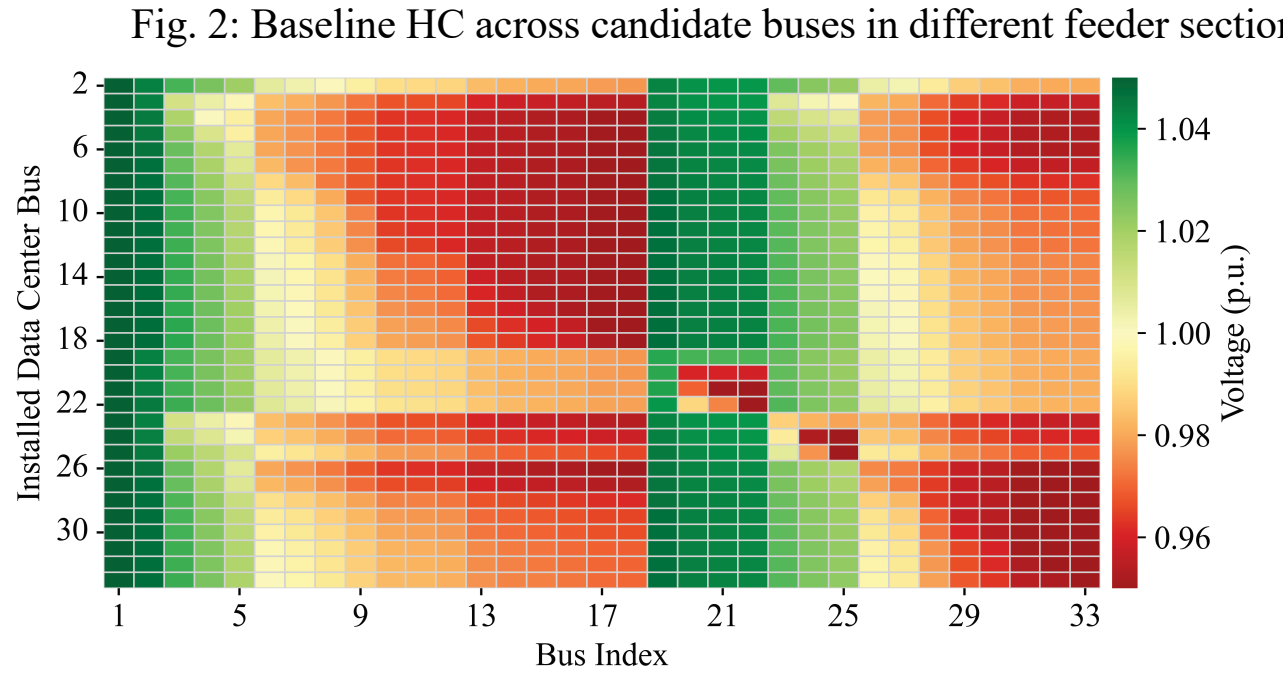

Fig. 3: Voltage profiles under baseline hosting capacity scenarios.

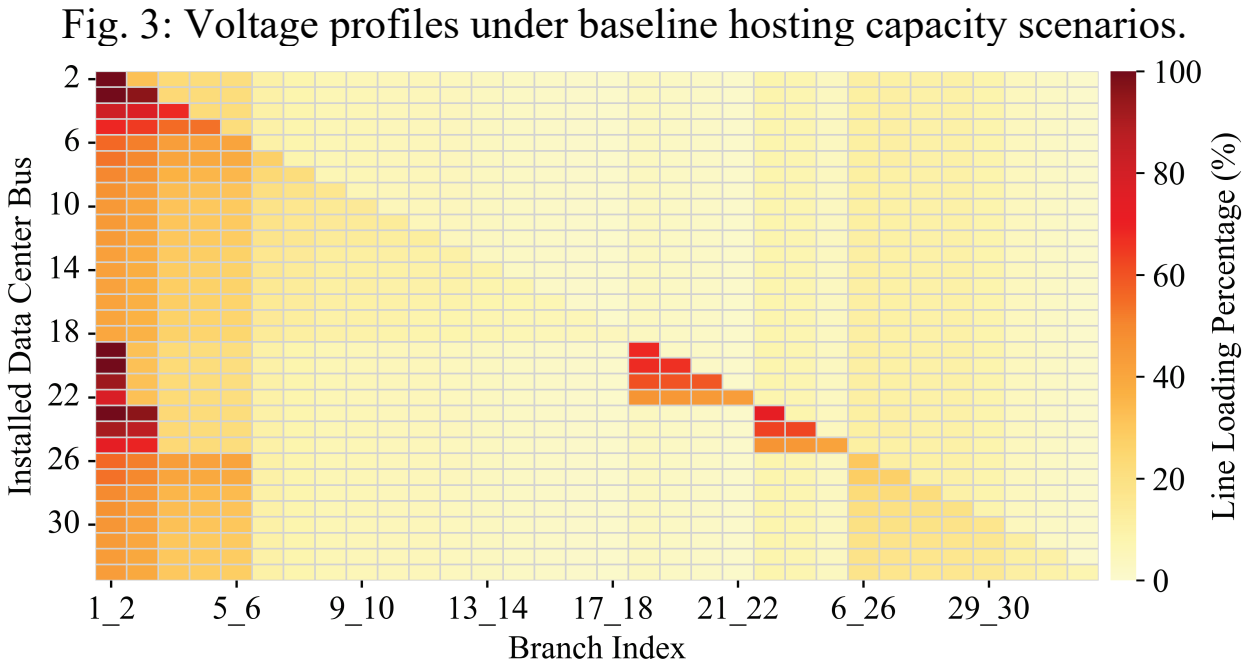

Fig. 4: Branch loading percentage under baseline hosting capacity scenarios.

Fig. 4 presents the corresponding branch loading percentage under the same baseline HC scenarios. Darker red regions indicate branches approaching their ampacity limits, which helps identify whether the HC is mainly constrained by thermal loading rather than voltage drop.

Based on the joint interpretation of these two heatmaps, the hosting capacity limitation mechanism of each candidate DC bus can be qualitatively classified into voltage-constrained, current-constrained, and mixed-constrained categories. Buses 2-3 and 19-20 are mainly current constrained, Buses 6-18 and 26-33 are mainly voltage constrained, while Buses 4-5 and 21-25 are mixed constrained because both voltage depression and branch loading are observed.

## IV. HC Enhancement with Flexible Resources

This section evaluates the improvement of DC hosting capacity enabled by flexible resources. Three types of resources are considered, including BESS, DDG, and STATCOM. The HC improvement ratio is calculated by comparing the enhanced HC with the original HC obtained in Section III. Since different candidate buses are limited by different operating constraints, the results are analyzed to reveal how active power support, sustained local generation, and reactive power compensation affect these constrained buses.

### *A. BESS-Based HC Improvement*

Fig. 5 demonstrates the HC improvement ratio when the BESS duration is fixed at 2 hours and the power sizing coefficient $\lambda^B$ is varied from 20% to 100%. The results show that increasing $\lambda^B$ leads to a clear increase in HC improvement across most candidate buses. This indicates that the BESS power rating has a dominant impact on HC enhancement because it directly determines the instantaneous active and reactive support capability available during limiting operating conditions. The improvement is more significant at electrically weak buses, while buses whose HC is mainly limited by branch ampacity show a more limited response.

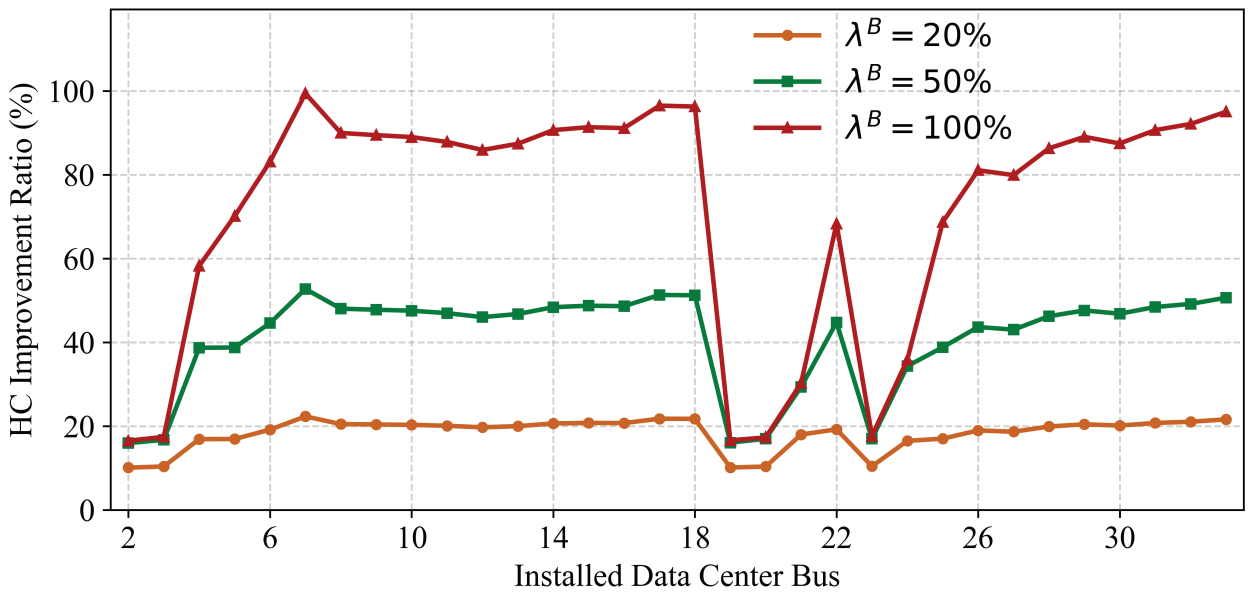

Fig. 5: Impact of BESS power sizing coefficient on HC improvement under fixed 2-h duration.

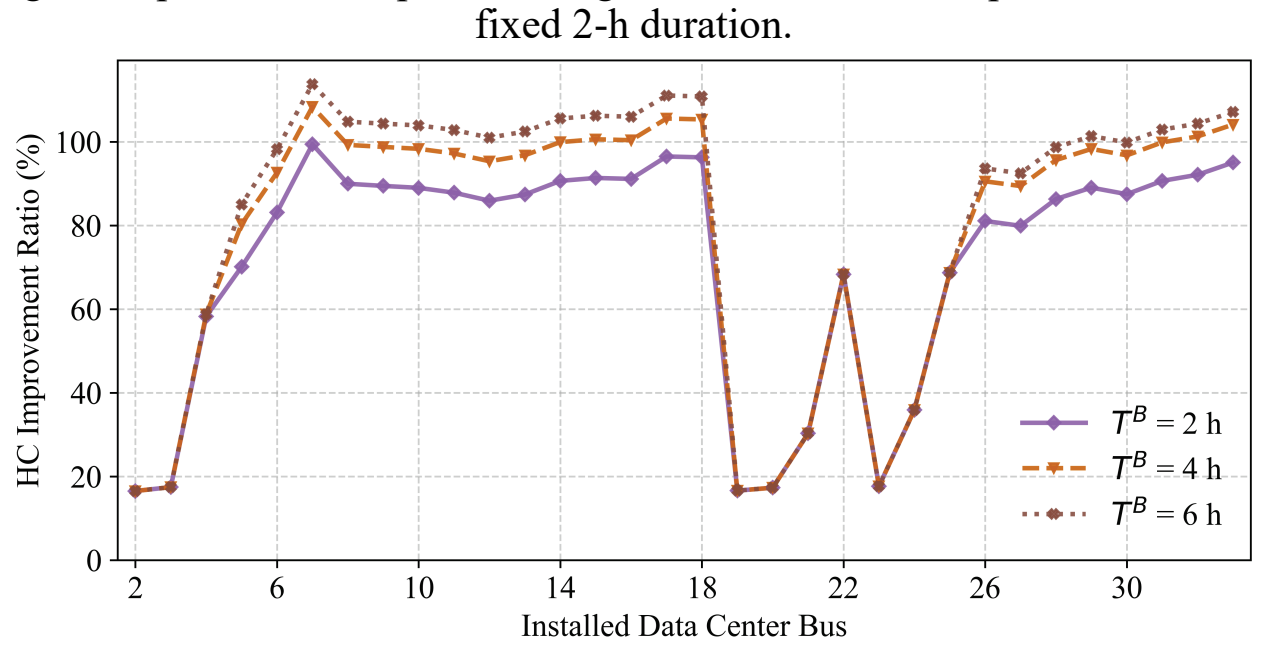

Fig. 6: Impact of BESS energy duration on HC improvement under fixed 100% power sizing.

Fig. 6 compares the HC improvement ratio for different BESS durations when the power sizing coefficient is fixed at 100%. Compared with the effect of $\lambda^B$, increasing $T^B$ from 2 to 6 hours provides a relatively smaller additional improvement. This suggests that the limiting factor is more related to instantaneous power support than to long-duration energy availability. Therefore, the BESS power rating is more influential than the energy duration in improving DC HC.

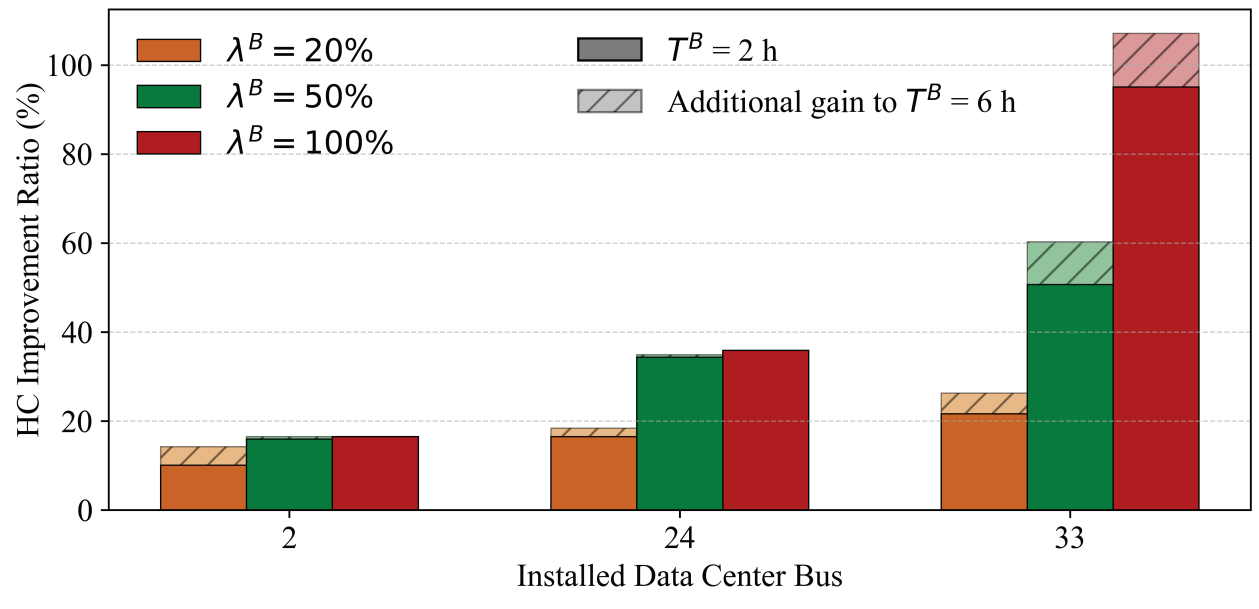

Fig. 7: Representative-bus comparison of BESS-induced HC improvement and duration marginal benefit.

Fig. 7 further compares three representative buses with different limiting mechanisms. Bus 2 represents a current-constrained location, Bus 24 represents a mixed-constrained location, and Bus 33 represents a voltage-constrained location. The results confirm that increasing $\lambda^B$ produces a much larger HC improvement than extending the energy duration. The hatched segments are relatively small compared with the solid bars, indicating that the marginal benefit of increasing $T^B$ is limited. Bus 33 shows the largest relative improvement because its baseline HC is low and voltage support from the BESS inverter can effectively relieve the voltage drop constraint. In contrast, Bus 2 shows a smaller improvement because its HC is mainly restricted by branch current loading.

*B. DDG-Based HC Improvement*

Fig. 8 shows the HC improvement ratio achieved by DDG under different sizing coefficients. As $\lambda^G$ increases from 20% to 100%, the HC improvement ratio increases consistently across candidate buses, indicating that local dispatchable active power support can effectively increase the allowable data center load. Since the DDG rating is determined proportionally to the original HC at each bus, the improvement curves remain relatively flat across buses, with only moderate variations caused by different voltage and branch loading constraints.

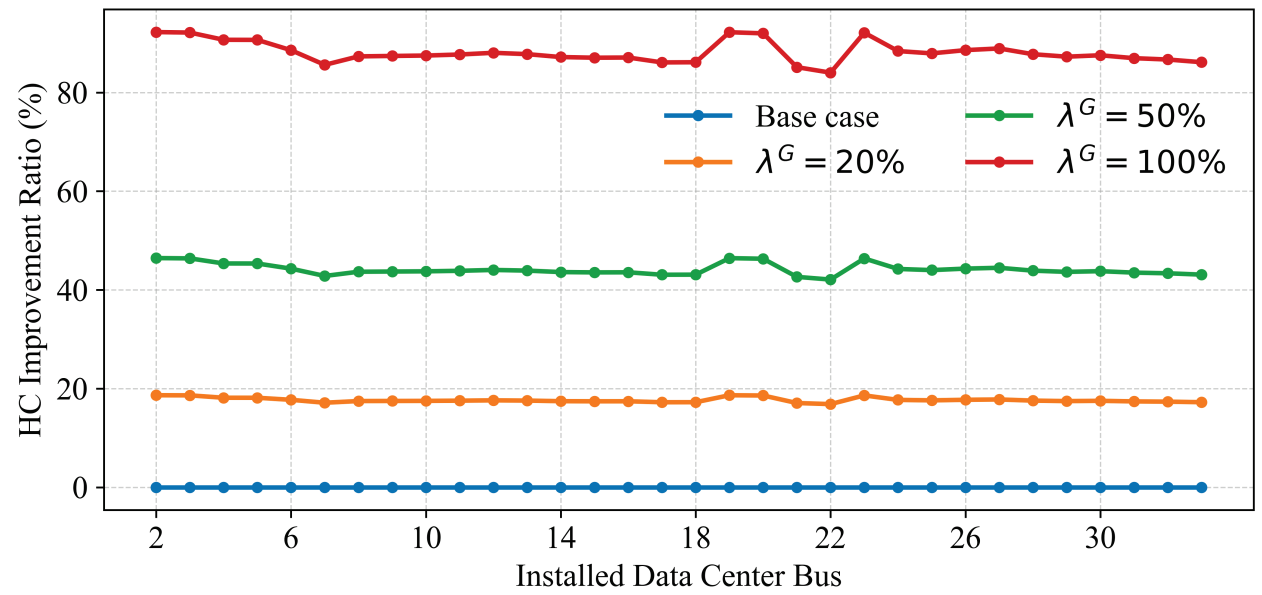

Fig. 8: DC HC improvement under different DDG sizing coefficients.

The results show that DDG provides a nearly proportional HC enhancement because it directly offsets part of the data center active power demand. However, the improvement ratio is still below the ideal proportional value in some locations, since voltage drops, branch losses, and ampacity constraints cannot be fully eliminated by active power support alone.

*C. STATCOM-Based HC Improvement*

Fig. 9 illustrates the HC improvement ratio achieved by STATCOM with different reactive power ratings from 1 to 3 MVAr. The results show that STATCOM provides significant improvement for buses whose baseline HC is mainly limited by voltage drop. In particular, the improvement is much higher for downstream buses, where local voltage support directly relaxes the voltage constraint. In contrast, buses close to the substation show limited improvement because their HC is more strongly affected by branch current loading rather than voltage drop.

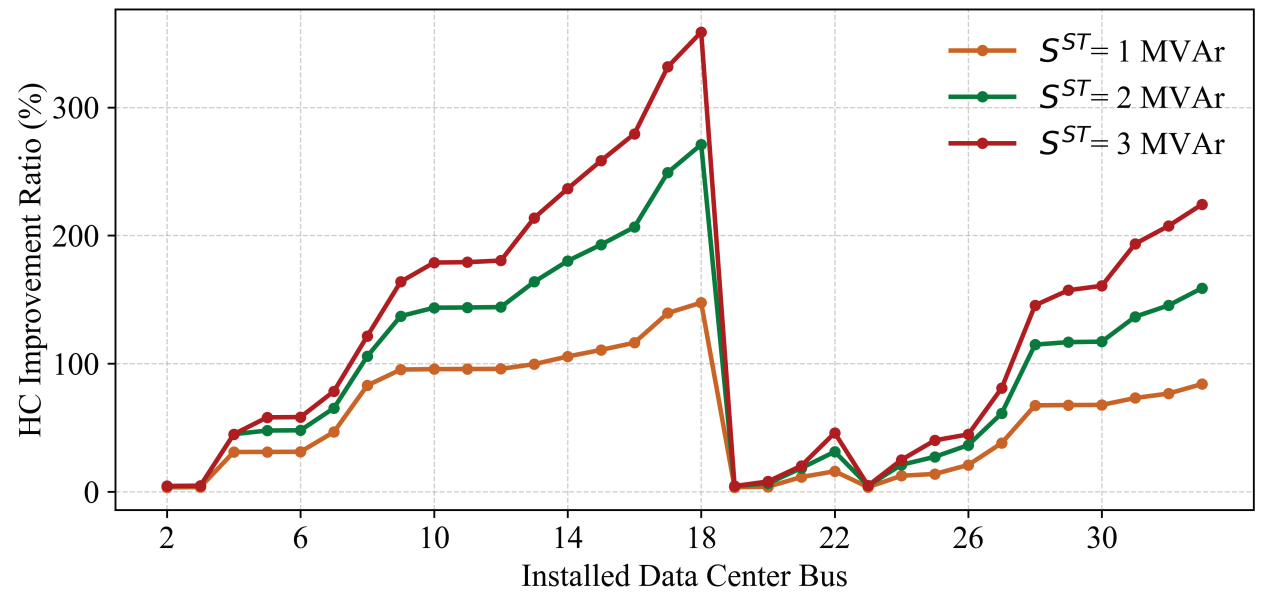

Fig. 9: Impact of STATCOM reactive support on hosting capacity.

Fig. 10 compares two STATCOM placement strategies for three representative buses. The co-located strategy places STATCOM at the DC installation bus, while the weakest-bus strategy places STATCOM at the voltage-critical bus identified after installing the DC load. Bus 2 represents a current-constrained case, Bus 24 represents a mixed-constrained case, and Bus 33 represents a voltage-constrained case.

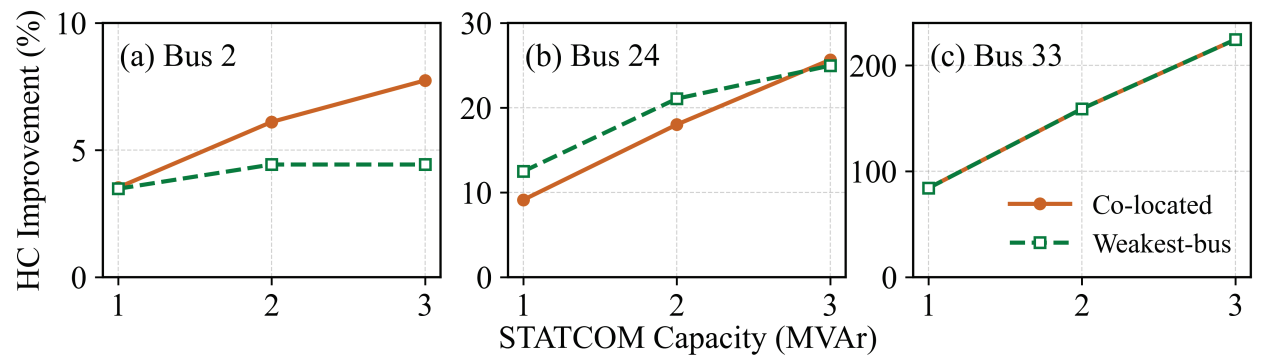

Fig. 10: Comparison of STATCOM placement strategies.

For Bus 2, placing STATCOM at the weakest bus at Bus 18 does not provide a clear advantage over the co-located placement because the HC is mainly limited by current loading. For Bus 24, the weakest-bus placement at Bus 25 provides higher improvement under 1 MVAr and 2 MVAr ratings, indicating that voltage support near the local weak point is beneficial for mixed-constrained cases. For Bus 33, the two strategies produce identical results because the DC bus itself is also the weakest voltage bus.

Table I compares the baseline HC with four enhancement scenarios under fixed resource settings: $\lambda^B = \lambda^G = 50\%$, $T^B = 2\text{h}$, and a 2-MVAr STATCOM. The BESS, DDG, and STATCOM cases represent individual resource deployment, while the combined case considers all three resources together in the active and reactive power balance equations. The result in Table I shows that all flexible resource scenarios improve DC hosting capacity, with the combined deployment achieving the largest overall enhancement by increasing the total HC from 91.89 MW to 194.46 MW. Among individual resources, STATCOM provides the strongest improvement in the minimum HC, indicating its effectiveness for weak voltage-constrained buses, while DDG achieves the highest maximum HC gain due to its direct active power support.

TABLE I. SUMMARY OF DC HOSTING CAPACITY IMPROVEMENT UNDER DIFFERENT FLEXIBLE RESOURCE SCENARIOS.

| Scenario | Min HC (MW) | Min Gain (%) | Max HC (MW) | Max Gain (%) | HC Sum (MW) | Total Gain (%) |
|---|---|---|---|---|---|---|
| Base Case | 0.425 | - | 7.953 | - | 91.89 | - |
| BESS | 0.643 | +51.2% | 9.224 | +16.0% | 120.43 | +31.1% |
| DDG | 0.608 | +43.1% | 11.647 | **+46.4%** | 133.14 | +44.9% |
| STATCOM | 1.578 | **+271.1%** | 8.305 | +4.4% | 129.09 | +40.5% |
| Combined | 1.929 | +353.9% | 13.212 | +66.1% | 194.46 | **+111.6%** |

## V. CONCLUSION

This paper evaluated the DC hosting capacity of a distribution network and analyzed the effectiveness of BESS, DDG, and STATCOM in improving the allowable DC load. The baseline results show that HC varies significantly across candidate buses due to different electrical locations and limiting constraints, including branch current loading, voltage drop, and mixed constraints. The resource sensitivity results indicate that BESS improves HC through instantaneous power support and inverter support, extending duration provides limited marginal gain. DDG provides relatively consistent improvement by directly offsetting part of the DC active power demand, while STATCOM is most effective for voltage-constrained buses through reactive power compensation. Under the selected settings, the combined deployment of BESS, DDG, and STATCOM achieves the highest overall HC improvement, demonstrating the complementary roles of active power support and reactive compensation.

## ACKNOWLEDGMENT

This This research was supported in part by the University of Houston Energy Transition Institute Seed Grant Program.

## REFERENCES

[1] Haoxiang Wan, Linhan Fang, and Xingpeng Li, "Grid Operational Benefit Analysis of Data Center Spatial Flexibility: Congestion Relief, Renewable Energy Curtailment Reduction, and Cost Saving", *IEEE PES General Meeting*, Montreal, Quebec, Canada, Jul. 2026.

[2] K. Lee, P. Zhao, A. Bhattacharya, B. K. Mallick and L. Xie, "An Active Learning-Based Approach for Hosting Capacity Analysis in Distribution Systems," in *IEEE Transactions on Smart Grid*, vol. 15, no. 1, pp. 617-626, Jan. 2024.

[3] Linhan Fang and Xingpeng Li, "Optimal BESS Sizing and Placement for Mitigating EV-Induced Voltage Violations: A Scalable Spatio-Temporal Adaptive Targeting Strategy", *arXiv*, Oct. 2025.

[4] J. Jian et al., "Supply Restoration of Data Centers in Flexible Distribution Networks With Spatial-Temporal Regulation," in IEEE Transactions on Smart Grid, vol. 15, no. 1, pp. 340-354, Jan. 2024.

[5] Jin Lu, Linhan Fang, Fan Jiang, and Xingpeng Li, "A Black Start Strategy for Hydrogen-integrated Renewable Grids with Energy Storage Systems", *IEEE International Conference on Energy Technologies for Future Grids*, Wollongon, Australia, Dec. 2025.

[6] Qiu, Ruixuan, et al. "Characterizing the Dynamic Hosting Capacity of Distribution Networks Integrated with Distributed GAI Data Center." IEEE Transactions on Industry Applications (2026).

[7] Rida Fatima, Linhan Fang, and Xingpeng Li, "A Reliability-Cost Optimization Framework for EV and DER Integration in Standard and Reconfigurable Distribution Network Topologies", *arXiv*, Nov. 2025.

[8] M. Rylander, J. Smith and W. Sunderman, "Streamlined Method for Determining Distribution System Hosting Capacity," in IEEE Transactions on Industry Applications, vol. 52, no. 1, pp. 105-111, Jan.-Feb. 2016.

[9] R. Lamedica, A. Geri, F. M. Gatta, S. Sangiovanni, M. Maccioni and A. Ruvio, "Integrating Electric Vehicles in Microgrids: Overview on Hosting Capacity and New Controls," in IEEE Transactions on Industry Applications, vol. 55, no. 6, pp. 7338-7346, Nov.-Dec. 2019.

[10] Linhan Fang, Elias Raffoul, and Xingpeng Li, "Diagnosis-Driven Co-planning of Network Reinforcement and BESS for Distribution Grid with High Penetration of Electric Vehicles", *arXiv*, Feb. 2026.

[11] F. Capitanescu, L. F. Ochoa, H. Margossian and N. D. Hatziargyriou, "Assessing the Potential of Network Reconfiguration to Improve Distributed Generation Hosting Capacity in Active Distribution Systems," in *IEEE Transactions on Power Systems*, vol. 30, no. 1, pp. 346-356, Jan. 2015.

[12] S. Wang, S. Chen, L. Ge and L. Wu, "Distributed Generation Hosting Capacity Evaluation for Distribution Systems Considering the Robust Optimal Operation of OLTC and SVC," in *IEEE Transactions on Sustainable Energy*, vol. 7, no. 3, pp. 1111-1123, July 2016.

[13] C. N. Acosta Campas, M. Madrigal Martinez and H. F. Ruiz Paredes, "PV Power Curtailment and BESS Management for Distribution Networks: A Practical Approach," in *IEEE Latin America Transactions*, vol. 21, no. 1, pp. 133-141, Jan. 2023.

[14] J. M. Home-Ortiz, O. D. Melgar-Dominguez, M. S. Javadi, J. R. S. Mantovani and J. P. S. Catalão, "Improvement of the Distribution Systems Resilience via Operational Resources and Demand Response," in *IEEE Transactions on Industry Applications*, vol. 58, no. 5, pp. 5966-5976, Sept.-Oct. 2022, doi: 10.1109/TIA.2022.3190241.

[15] H. Akagi, S. Inoue and T. Yoshii, "Control and Performance of a Transformerless Cascade PWM STATCOM With Star Configuration," in *IEEE Transactions on Industry Applications*, vol. 43, no. 4, pp. 1041-1049, July-aug. 2007, doi: 10.1109/TIA.2007.900487.